\documentclass{article}
\usepackage{ijcai23}

\usepackage{times}
\usepackage{soul}
\usepackage{url}
\usepackage[hidelinks]{hyperref}
\usepackage[utf8]{inputenc}
\usepackage[small]{caption}
\usepackage{graphicx}
\usepackage{amsmath}
\usepackage{amsthm}
\usepackage{amssymb}
\usepackage{bm}
\usepackage{booktabs}
\usepackage{color}
\usepackage{algorithm}
\usepackage{algorithmic}
\usepackage{multirow}
\usepackage{hhline}
\usepackage[switch]{lineno}

\newtheorem{example}{Example}
\newtheorem{theorem}{Theorem}
\newtheorem{definition}{Definition}
\newtheorem{problem}{Problem}
\newtheorem{remark}{Remark}
\newtheorem{corollary}{Corollary}
\newtheorem{proposition}{Proposition}
\newcommand{\oomit}[1]{}

\title{Provable Reach-avoid Controllers Synthesis Based on Inner-approximating Controlled Reach-avoid Sets}

\author{
    Jianqiang Ding$^1$, Taoran Wu$^{1,2}$, Yuping Qian$^3$, Lijun Zhang$^{1,2}$ and Bai Xue$^{1,2}$ 
    \affiliations
    \textsuperscript{\rm 1}State Key Lab. of Computer Science, Institute of Software, CAS, Beijing, China \\
    \textsuperscript{\rm 2} University of Chinese Academy of Sciences, Beijing, China \\
    \textsuperscript{\rm 3}Tsinghua University, Beijing, China
    \emails
    dingjianqiang0x@gmail.com,
    qianyuping@tsinghua.edu.cn,
    \{wutr,zhanglj,xuebai\}@ios.ac.cn
}

\begin{document}

\maketitle

\begin{abstract}
In this paper, we propose an approach for synthesizing provable reach-avoid controllers, which drive a deterministic system operating in an unknown environment to safely reach a desired target set. The approach falls within the reachability analysis framework and is based on the computation of inner-approximations of controlled reach-avoid sets (CRSs). Given a target set and a safe set, the controlled reach-avoid set is the set of states such that starting from each of them, there exists at least one controller to ensure that the system can enter the target set while staying inside the safe set before the target hitting time. Therefore, the boundary of the controlled reach-avoid set acts as a barrier, which separating states capable of achieving the reach-avoid objective from those that are not, and thus the computed inner-approximation provides a viable space for the system to achieve the reach-avoid objective. Our approach for synthesizing reach-avoid controllers mainly consists of three steps. We first learn a safe set of states in the unknown environment from sensor measurements based on a support vector machine approach. Then, based on the learned safe set and target set, we compute an inner-approximation of the CRS. Finally, we synthesize controllers online to ensure that the system will reach the target set by evolving inside the computed inner-approximation. The proposed method is demonstrated on a Dubin's car system. 
\end{abstract}

\section{Introduction}
Safety-critical systems are becoming ubiquitous and a basic part of our life. To deploy safety-critical systems, it is of vital importance to assure their safety, i.e., their states should satisfy some safety constraints and thus evolve within a safe set \cite{amodei2016concrete}. These safety constraints might be caused either internally by physical limitations such as actual saturation or by external factors such as surrounding obstacles to be avoided. 
Satisfaction of these safety constraints is crucial and should be considered seriously during the control design phase because their violation can lead to catastrophic consequences.

Safety-critical systems such as autonomous vehicles, industrial robots, and multi-robot systems are often deployed in uncertain and complex environments, which are required to respect safety-critical constraints while advancing a given task. When operating in unknown and dynamic environments with insufficient advanced information regarding workspace, controllers which translate sensory information from the environment into safe control actions are of paramount importance. In recent years, the development of machine learning algorithms such as supervised learning and (safe) reinforcement learning  have created unprecedented opportunities to control modern systems \cite{6287086,pmlr-v70-achiam17a}. However, machine learning also poses great challenges for developing high-assurance systems.  While many learning-based approaches have  been proposed to train controllers to accomplish complex tasks with improved empirical performance, the lack of safety certificates for the learning-enabled components has been a fundamental hurdle that blocks the massive deployment of the learned solutions. For decades, mathematical control certificates  such as control Lyapunov functions \cite{2021Lyapunov}, control contraction metrics \cite{7852456}, and control barrier functions \cite{ames2019control} have been developed as proofs that the desired properties of the system are satisfied in closed-loop with certain control policies. Among them, control barrier functions, whose certain superlevel sets form safe sets, are mainly used to find an action driving systems to stay within a safe set for all the time. With such functions, there are a few approaches to guarantee safety in the machine learning framework. For instance, an offline controller synthesis framework, which integrates existing model-free reinforcement learning algorithms with control barrier functions, was proposed for guaranteeing safety in \cite{cheng2019}. 

Besides the safety objective, safety-critical systems often require simultaneous satisfaction of multiple performance specifications. In order to simultaneously achieve safety and stability performances (i.e., stabilize the system into an equilibrium state safely), many control design methods have been proposed in the literature, which integrate control Lyapunov functions with control barrier functions.  These methods have been successfully designed for a broad range of applications such as adaptive cruise control \cite{xu2017correctness} and  safe control of robots \cite{agrawal2017discrete}. However, when these two objectives were in conflict, no feedback controllers can be designed. To deal with conflicting safety and stabilization objectives, an optimization based controller was developed in \cite{7782377} such that safety is strictly guaranteed while convergence to goal is relaxed. Besides, these approaches typically assume that a valid control barrier function is provided. Nevertheless, this assumption does not hold, especially for systems operating in uncertain environments for which they have no knowledge of surrounding unsafe factors. 
\cite{Fabio2016,8794118} synthesized occupancy map functions for navigation and planning purposes based on the use of kernel machines \cite{cristianini_shawe-taylor_2000}.  In \cite{9341190}, the authors parameterized a control barrier function by a SVM, and used a supervised learning approach to characterize regions of the state-space as safe or unsafe based on collected data.  
 \cite{8967981} proposed a method which incrementally learns a linear control barrier function by clustering expert demonstrations into linear subspaces and fitting low dimensional representations. However, in these cases learning a control barrier function is decoupled from finding a policy such that the obtained barrier function may be no longer valid, which cannot assure the existence of control actions enforcing safety objectives. Consequently, \cite{9303785} proposed an approach to learn control barrier functions using expert demonstrations of safe trajectories. 

In contrast to the aforementioned methods, this paper investigates the reach-avoid controllers synthesis problem, which formulates many important engineering problems such as collision avoidance as well as target surveillance. The problem is to synthesize a controller such that the system operating in an unknown environment satisfies the objective of joint safety and reachability, i.e., reach a desired target destination safely. In our approach we also learn a safe set via sensor measurements. Further, in order to assure existence of reach-avoid controllers, we find a subset of the learned safe set, i.e., an inner-approximation of the controlled reach-avoid set (CRS). This set carves out a viability space for the system to be capable of achieving the reach-avoid objective in the unknown environment. On the other hand, with known environments, there are some studies on synthesizing reach-avoid controllers offline such as \cite{fan2021controller,wang2021verification,kochdumper2021aroc}, which generally rely on set propagation approaches and computing over-approximations (i.e., super-sets) of reach states \cite{casagrande2022parameter}. However, due to the wrapping effect in computing over-approximations (i.e., the accumulation of approximation errors over chains of successive
time steps), overly pessimistic over-approximations often render many reach-avoid controller synthesis problems unsolved, especially for complex real-world applications with large time horizons. Thus, the target hitting time in these studies has to be deterministic. In contrast, in this work it is uncertain.

%


In this paper we propose a framework for synthesizing provable reach-avoid controllers for deterministic systems operating in unknown environments. 
Within this framework, the computation of inner-approximations of the CRS plays a fundamental role, as the existence of reach-avoid controllers is guaranteed when the system operates within the computed inner-approximation. Our approach for synthesizing reach-avoid controllers mainly consists of three steps. The first step is to classify safe and unsafe states, and learn a safe set from sensor measurements based on a SVM. The system should be controlled to operate within this safe set in order to avoid dangerous situations such as collision with obstacles. However, we cannot guarantee the existence of reach-avoid controllers for every state in this learned safe set. Consequently, we further classify a set of states, for which the existence of reach-avoid controllers is ensured. For this sake, we will compute an inner-approximation of the CRS, which is a complicated nonlinear problem arising in dynamical systems and control theory. Although there are some works on computing outer-approximations of CRSs \cite{han2018controller,zhaoCDC}, efficient methods to compute its counterpart(i.e., inner-approximations) are rare. In this work we derive a set of new constraints for computing inner-approximations of the CRS. This set of constraints is convex and thus is possible to generate an inner-approximation of CRSs efficiently using convex optimization. It is constructed from a probabilistic perspective and is inspired by the work \cite{xue2021reach}, which studies the problem of inner-approximating the reach-avoid set in the probabilistic sense. The $0$-reach-avoid set in \cite{xue2021reach}, which is a set of all states such that the system driven by random (control/disturbance) signals is able to  enter a  target set safely with a probability being larger than zero, is an inner-approximation of the CRS when random signals are controllers. Based on the computed inner-approximation, we synthesize a reach-avoid controller online. The same steps can be repeated based on updated sensor measurements for reaching new target sets. Finally, we demonstrate our proposed approach on a Dubin's car system.


The contributions of this work are summarized below. 
\begin{enumerate}
    \item A framework for synthesizing provable controllers, which enforces the reach-avoid objective for deterministic system operating in an unknown environment, is presented. The synthesis of provable reach-avoid controllers is built upon the computation of CRSs.
    \item A set of new convex constraints is derived for inner-approximating CRSs of deterministic systems from a probabilistic perspective.  
    When the datum involved are polynomials, i.e., the system has polynomial dynamics in the state variables, and the safe set and target set are semi-algebraic sets, the gain of these constraints facilitates the computation of CRSs by solving convex optimization. 
   Moreover, some practical concerns can also be addressed via solving this set of constraints. For instance, compared with one in \cite{xue2021reach}, this set of constraints is simpler, which could be solved more efficiently; an upper bound for the minimal target hitting time, which can be used to verify whether the target hit occurs within some tolerable time limit, can be obtained.
\end{enumerate}


The following  notations will be used throughout the rest of this paper: $\mathbb{R}^n$ and $\mathbb{N}$ denote
the set of n-dimensional real vectors and non-negative integers, respectively; the closure of a set $\mathcal{X}$ is denoted by $\overline{\mathcal{X}}$, and the boundary by $\partial \mathcal{X}$; vectors are denoted by boldface letters; $\mathbb{R}[\cdot]$ denotes the ring of polynomials in variables given by the argument; $\sum[\bm{x}]$ denotes the set of sum of squares polynomials, i.e.,
\[\sum[\bm{x}]=\left \{p(\bm{x})\middle |\;
\begin{aligned}
 &p(\bm{x})=\sum_{i=1}^k q_i^2(\bm{x}), 
 &q_i(\bm{x})\in \mathbb{R}[\bm{x}]
\end{aligned}
\right\}.
\]
\section{Preliminaries}
\label{sec:pre}
In this section we introduce the system, reach-avoid controller synthesis problem of interest, and $0$-reach-avoid sets. 

\subsection{Problem Formulation}
Consider a discrete-time control system, whose dynamics are described by
\begin{equation}
\label{system}
    \bm{x}(k+1)=\bm{f}(\bm{x}(k),\bm{u}(k)),
\end{equation}
where $\bm{x}(\cdot):\mathbb{N}\rightarrow \mathcal{D}$ is the state with $\mathcal{D}\subseteq \mathbb{R}^n$, and $\bm{u}(\cdot): \mathbb{N}\rightarrow \mathcal{U}$ with $\mathcal{U}\subseteq \mathbb{R}^m$ is the control input, 
and $\bm{f}(\cdot,\cdot): \mathbb{R}^n \times \mathbb{R}^m \rightarrow \mathbb{R}^n$ is continuous with respect to the arguments.

The evolution of system \eqref{system} is driven by a controller, which is defined below. 
\begin{definition}
    A controller $\pi$ for system \eqref{system} is a sequence $(\bm{u}(l))_{l\in \mathbb{N}}$, where $\bm{u}(\cdot):\mathbb{N}\rightarrow \mathcal{U}$. We define $\mathbb{U}$ as the set of all controllers.  
\end{definition}

    Given a controller $\pi\in \mathbb{U}$ and an initial state $\bm{x}_0\in \mathcal{D}$, we can obtain a trajectory keeping tracking of the evolution of system \eqref{system}.

    \begin{definition}
        Given an initial state $\bm{x}_0\in \mathbb{R}^n$ and a controller $\pi=(\bm{u}(l))_{l\in \mathbb{N}}$, the trajectory of system \eqref{system}, induced by $\bm{x}_0$ and $\pi$, is a sequence $(\bm{\phi}_{\bm{x}_0}^{\pi}(l))_{l\in \mathbb{N}}$ satisfying  
        \[
        \begin{cases}
        &\bm{\phi}_{\bm{x}_0}^{\pi}(l+1)=\bm{f}(\bm{\phi}_{\bm{x}_0}^{\pi}(l),\bm{u}(l)),\forall l\in \mathbb{N}\\
        &\bm{\phi}_{\bm{x}_0}^{\pi}(0)=\bm{x}_0.
        \end{cases}\]
    \end{definition}

    Given a target set $\mathcal{X}_r\subseteq \mathbb{R}^n$ and a safe set $\mathcal{C}\subseteq \mathcal{D}$, the CRS is defined below. 
    \begin{definition}
        The {\em CRS} $\mathcal{R}$ for system \eqref{system} is the set of all initial states $\bm{x}_0\in \mathcal{C}$ such that starting from each of them, there exists at least one controller $\pi\in \mathbb{U}$ to ensure that the resulting trajectory can hit the target set $\mathcal{X}_r$ in a finite time $k\in \mathbb{N}$ while staying inside the safe set $\mathcal{C}$ before $k$, i.e., 
        \begin{equation*}
\mathcal{R}= \left \{\bm{x}_0\in \mathcal{C} \middle |\;
\begin{aligned}
&\exists \pi \in \mathbb{U}. \exists k\in \mathbb{N}.\\
&[ \bm{\phi}_{\bm{x}_0}^{\pi}(k)\in \mathcal{X}_r \wedge \bigwedge_{l=0}^k \bm{\phi}_{\bm{x}_0}^{\pi}(l)\in \mathcal{C}]
  \end{aligned}
  \right\}.
\end{equation*}       
\end{definition}

It is observed that only when the state of system \eqref{system} evolves within the CRS, we can guarantee the existence of controllers such that system \eqref{system} achieves the reach-avoid objective. Generally, it is challenging to compute the exact CRS. Consequently, its outer- or inner-approximations are often resorted to in practice, especially for formally reasoning about properties of systems. In order to guarantee the existence of controllers enforcing the reach-avoid objective, we in this paper compute inner-approximations for solving the reach-avoid controllers synthesis problem of interest.

\begin{problem}[Reach-avoid Controllers Synthesis]
Suppose that system \eqref{system} operates in an unknown environment and dangerous states can be detected via sensors onboard. Given a set $\mathcal{X}_r$ of target states and an initial state $\bm{x}_0\in \mathcal{D}$, we attempt to synthesize a controller $\pi \in \mathbb{U}$ such that system \eqref{system} is able to reach the target set $\mathcal{X}_r$ while avoiding all of dangerous states encountered in the operating process. 
\end{problem}

In this paper we assume that the system dynamics are certain. Our approach for synthesizing reach-avoid controllers is also applicable to systems with uncertain dynamics. When uncertain dynamics are present, the uncertain dynamics can be learned using Gaussian process with measurements collected online by sampling from the system, as done in \cite{berkenkamp2017safe,cheng2019}. Then, confidence intervals, defined based on the learned system dynamics model, are used to synthesize reach-avoid controllers. 

\subsection{\texorpdfstring{$0$-Reach-avoid Sets}{}}
In our approach we will inner-approximate the CRS $\mathcal{R}$ based on the computation of an inner-approximation of $0$-reach-avoid sets in the probabilistic setting \cite{xue2021reach}. Therefore, in this subsection we will give an introduction on the $0$-reach-avoid set. 

We endow the set $\mathcal{U}$ with an appropriately arbitrary but fixed probability measure $P$, and suppose that the random vectors, $\bm{u}(0)$, $\bm{u}(1)$, $\ldots$, are independent and identically distributed (i.i.d), and take values in $\mathcal{U}$ with the following probability distribution, 
\[\begin{split}
&\text{Prob}(\bm{u}(l)\in \mathcal{A})=P(\mathcal{A}), \forall l\in \mathbb{N}, \mathcal{A}\subseteq \mathcal{U},\\
&P(\mathcal{U})=1.
\end{split}
\]
$E[\cdot]$ is the expectation induced by the probability distribution $P$. Then, given system \eqref{system} with the probability distribution $P$, a controller $\pi=(\bm{u}(i))_{i\in \mathbb{N}}$ is a stochastic process defined on the canonical sample space $\Omega=\mathcal{U}^{\infty}$, endowed with its product topology $\mathcal{B}(\Omega)$, with probability measure $P^{\infty}$. The probability measure $P^{\infty}$ is induced by the probability measure $P$, and its associated expectation is denoted by $E^{\infty}[\cdot]$.

\begin{definition}[\cite{xue2021reach}]
The $p$-reach-avoid set $\mathcal{R}_p$ is the set of all initial states that each gives rise to a set of trajectories which, with a probability being larger than $p\in [0,1)$, eventually enter the target set $\mathcal{X}_r$ while remaining inside the safe set $\mathcal{C}$ until the target hit, i.e., 
\begin{equation*}
\mathcal{R}_p=\left\{\bm{x}_0\in \mathcal{C} \middle|\; 
\begin{aligned}
P^{\infty}\Big(&\exists k\in \mathbb{N}. \bm{\phi}_{\pi}^{\bm{x}_0}(k)\in \mathcal{X}_r\bigwedge\\
&\forall l\in [0,k]\cap \mathbb{N}. \bm{\phi}_{\pi}^{\bm{x}_0}(l)\in \mathcal{C}\Big)> p
\end{aligned}
\right\}.
\end{equation*} 
\end{definition}

As commented in Remark 1 in \cite{xue2021reach}, the 0-reach-avoid set $\mathcal{R}_0$ is the set of all initial states that each gives rise to a set of trajectories which will enter $\mathcal{X}_r$ safely with a probability being larger than 0. That is, there exists a non-empty set of controllers $\pi\in \Pi$ such that system \eqref{system} originating from $\mathcal{R}_0$ will be driven to enter $\mathcal{X}_r$ safely. Thus, we in the paper will inner-approximate the CRS $\mathcal{R}$ via computing inner-approximations of the $0$-reach-avoid set.

\section{Provable Reach-avoid Controllers Synthesis}
\label{sec:CS}
\begin{figure}[htb!]
\center
\includegraphics[angle=-90,width=0.9\linewidth]{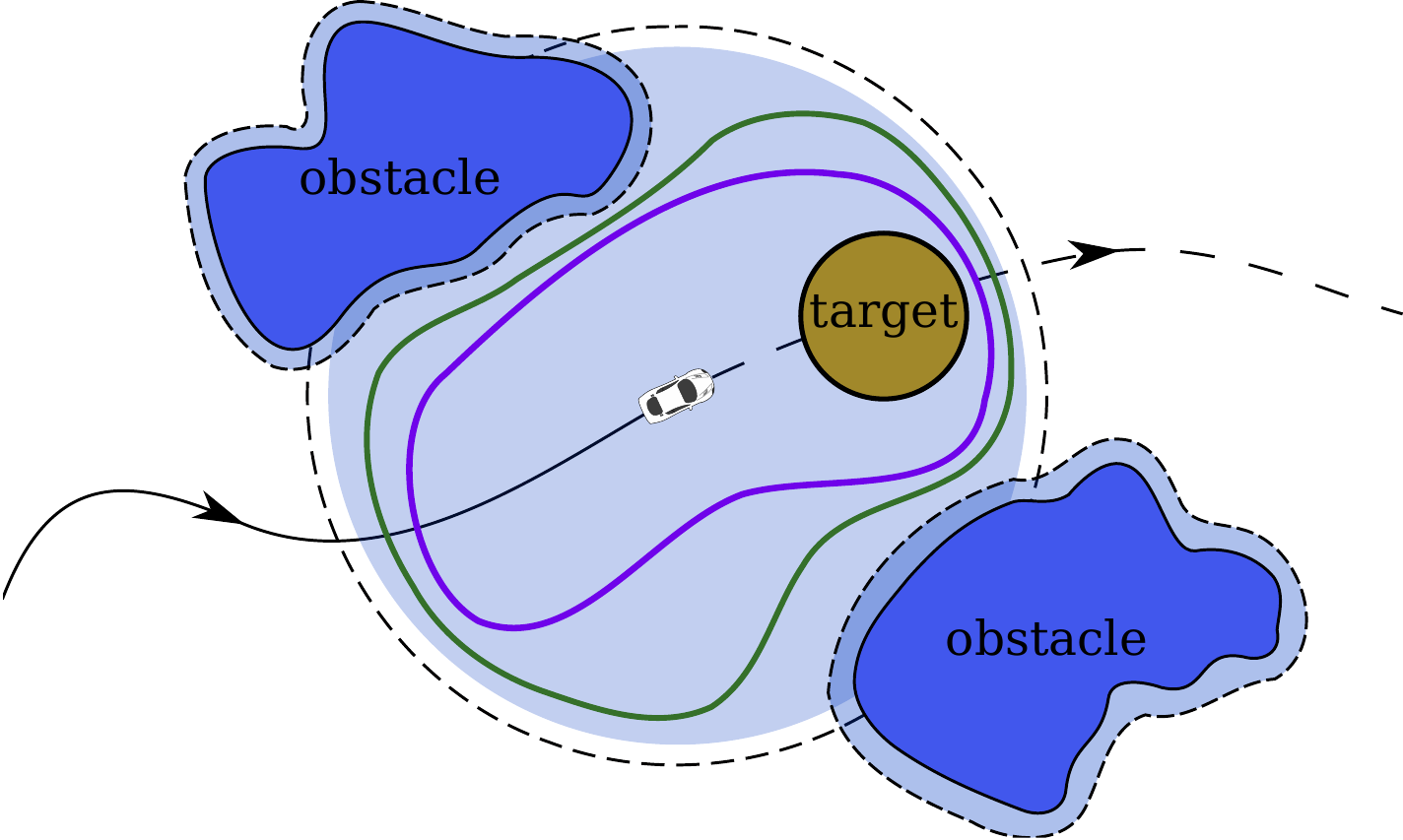} 
\caption{an illustration of our reach-avoid controllers synthesis method (It  begins with learning a safe set from the sensor measurements onboard, which is bounded by the green curve, to avoid collision with possible obstacles around, and then computes an inner-approximation of the CRS, which is bounded by the purple curve, in order to guarantee the existence of reach-avoid controllers. Finally, via constraining the system within the computed inner-approximation, we synthesize a reach-avoid controller online, guaranteeing the reach of the target set safely.)}
\label{pipeline}
\end{figure}

In this section we introduce our approach for synthesizing provable reach-avoid controllers for system \eqref{system} operating in unknown environments. The framework of our approach is presented in Fig. \ref{pipeline}.

\subsection{Learning Safe Sets} \label{subsec:LSS}
We use the supervised learning method in \cite{9341190} to learn a safe set based on collected datum from sensors onboard.

Consider system \eqref{system} evolving in $\mathcal{D}\subseteq \mathbb{R}^n$ and equipped with LiDAR sensors such as Veldyne that are able to provide depth information with high accuracy. By virtual of the depth measurement vector $\{\bm{z}_i\}_{1\leq i\leq N}$ at time $k$, where $N$ is the number of samples, system \eqref{system} can detect unsafe states and define the set of safe and unsafe samples. These samples will be used for training SVM classifiers and learning a safe set in which the system is allowed to operate.


The learning approach to be used for determining a safe set will be kernel SVMs \cite{cristianini_shawe-taylor_2000}. Suppose a dataset $\mathcal{T}=\{(\bm{x}_1,y_1),\ldots, (\bm{x}_N,y_N)\}$ is provided, where $\bm{x}_i\in \mathbb{R}^n$ is a point in the n-dimension space and $y\in \mathcal{Y}=\{-1,1\}$ is a label associated with the vector $\bm{x}_i$ for all $i\in \{1,\ldots,N\}$. If $y_i=1$, the sate $\bm{x}_i$ is safe; otherwise, the state $\bm{x}_i$ is classified as unsafe. Since the domain $\mathcal{D}$ consists of states which are either safe or unsafe, this separation can be cast as a binary SVM classification problem. The resulting nonlinear, biased-penalty SVM optimziation problem is presented below.
\begin{equation}
\label{SVM}
    \begin{split}
        &\min \frac{1}{2}\|\bm{w}\|_2^2+C^+ \sum_{i\mid y_i=1}^N \xi_i+C^- \sum_{j\mid y_j=-1}\xi_j\\ 
&\text{s.t.}\begin{cases}
    &y_i\cdot (\bm{w}^{\top}\bm{\phi}(\bm{x}_i)+b)\geq 1-\xi_i,\\
    &\xi_i\geq 0, \\
    &i=1,\ldots,N,
\end{cases}
\end{split}
\end{equation}
where $C^+, C^->0$ are constants penalizing misclassification of the positive and negative samples, and $\bm{\phi}(\cdot): \mathbb{R}^n\rightarrow \mathbb{R}^d$ is a nonlinear mapping into a higher dimensional space, $\bm{w}\in \mathbb{R}^d$ are coefficients, and $b\in \mathbb{R}$ is a bias term. In \eqref{SVM} there are two separate costs for the positive and negative classes. Unequal costs permit a greater bias towards correctly classifying one class over the other. In safety-critical scenarios, it is imperative that unsafe states can be classified unsafe, whereas all the safe states need not strictly be classified as safe. Therefore, we need a large $C^-$ (e.g., $10^{12}$) and a small $C^+$ in practice for classifying these  states.


After learning the set $\mathcal{C}=\{\bm{x}\in \mathcal{D}\mid 1-\bm{w}^{\top}\bm{\phi}(\bm{x})-\bm{b}\leq 0\}$ of states, a potential technique to synthesize a reach-avoid controller is model predictive control \cite{camacho2013model}, which is widely used in both academic and industry communities for synthesizing controllers such that the system satisfies multiple constraints. Given a finite time horizon from 0 to $N'$, the model predictive control for synthesizing a reach-avoid controller is formulated as follows. 

\begin{equation}
\label{MPC}
\begin{split}
    &\min_{\bm{u}(1),\ldots,\bm{u}(N'-1)} l(\bm{x}_1,\ldots,\bm{x}_{N'})\\
&\text{s.t.}
\begin{cases}
     &\bm{x}(k+1)=\bm{f}(\bm{x}_k,\bm{u}_k), k=0,\ldots,N'-1,\\
    &\bm{x}_k\in \mathcal{C}, k=0,\ldots,N', \\
    &\bm{x}_{N'}\in \mathcal{X}_r, \\
    &\bm{x}(0)=\bm{x}_0,
    \end{cases}
    \end{split}
\end{equation}
where $l(\bm{x}_1,\ldots,\bm{x}_{N'})$ is some performance to be optimized.

Due to uncertainties on the target hitting time, we cannot ensure the existence of control actions $\bm{u}(0),\ldots,\bm{u}(N'-1)$ satisfying the constraints in optimization \eqref{MPC}. Also, a larger  time horizon, i.e., $N'$ is larger, will increase the computation time in solving optimization \eqref{MPC}, thus degrading the real-time performance. Below we will address this challenge by computing an inner-approximation of the CRS and taking a simple sampling controller synthesis method. 

\subsection{Inner-approximating CRSs}
\label{subsec:IACRS}
 As mentioned previously, the problem of inner-approximating $\mathcal{R}$ can be transformed into a problem of inner-approximating the $0$-reach-avoid set $\mathcal{R}_0$. In this section we will derive a set of new constraints for inner-approximating $\mathcal{R}$, via inner-approximating $\mathcal{R}_0$.



 A set of constraints for inner-approximating $\mathcal{R}_0$ is formulated in Proposition \ref{pro1} below. Starting from the computed inner-approximation, there exists a controller such that system \eqref{system} can reach the target set eventually while staying inside {\em the safe set $\mathcal{C}$} before the first target hitting time.

\begin{proposition}[Corollary 2, \cite{xue2021reach}]
\label{pro1}
Suppose that the set $\widehat{\mathcal{C}}$ includes the set of states which system \eqref{system} starting from the safe set $\mathcal{C}$ visits within the first step, i.e., 
\begin{equation}
\label{sets}
    \widehat{\mathcal{C}}\supseteq \{\bm{x}\in \mathbb{R}^n\mid \bm{x}=\bm{f}(\bm{x}_0,\bm{u}),\bm{x}_0\in \mathcal{C}, \bm{u}\in \mathcal{U}\}\cup \mathcal{C}.
\end{equation}
 If there exist bounded functions $v(\bm{x}): \widehat{\mathcal{C}}\rightarrow \mathbb{R}$ and $w(\bm{x}):\widehat{\mathcal{C}}\rightarrow \mathbb{R}$ such that for $\bm{x}\in \widehat{\mathcal{C}}$, 
\begin{equation}
\label{constraint}
    \begin{cases}
    &E[v(\widehat{\bm{f}}(\bm{x},\bm{u}))]-v(\bm{x})\geq 0, \\
    &v(\bm{x})\leq 1_{\mathcal{X}_r}(\bm{x})+E[w(\widehat{\bm{f}}(\bm{x},\bm{u}))]-w(\bm{x}),
    \end{cases}
\end{equation}
then $\{\bm{x}\in \mathcal{C}\mid v(\bm{x})>0\}$ is an inner-approximation of the $0$-reach-avoid set $\mathcal{R}_0$, i.e., $\{\bm{x}\in \mathcal{C}\mid v(\bm{x})>0\}\subseteq \mathcal{R}_0$, where 
\[\widehat{\bm{f}}(\bm{x},\bm{u})=1_{\mathcal{C}\setminus \mathcal{X}_r}(\bm{x})\cdot\bm{f}(\bm{x},\bm{u})+1_{\mathcal{X}_r}(\bm{x})\cdot\bm{x}+1_{\widehat{\mathcal{C}}\setminus \mathcal{C}}(\bm{x})\cdot\bm{x}\]
with \[1_{\mathcal{A}}(\bm{x})=
\begin{cases}
    1, \text{if~}\bm{x}\in \mathcal{A},\\
    0,\text{otherwise}.
\end{cases}
\]
being the indicator function ($\mathcal{A} \in \{\mathcal{C}\setminus \mathcal{X}_r, \mathcal{X}_r,\widehat{\mathcal{C}}\setminus \mathcal{C}\}$). 
\end{proposition}

\begin{remark}
In \cite{xue2021reach}, the system 
\begin{equation}
\label{new_sys}
\begin{cases}
& \bm{\psi}_{\pi}^{\bm{x}_0}(i+1)=\widehat{\bm{f}}(\bm{\psi}_{\pi}^{\bm{x}_0}(i),\pi(i)), \forall i\in \mathbb{N},\\
&\bm{\psi}_{\pi}^{\bm{x}_0}(0)=\bm{x}_0.
\end{cases}
\end{equation}
is used, whose behaviors are the same with the ones of system \eqref{system} in $\mathcal{C}\setminus \mathcal{X}_r$. For system \eqref{system}, we have that $\forall k\in \mathbb{N}. \bm{\psi}_{\pi}^{\bm{x}_0}(k)\in \widehat{\mathcal{C}}$, and its $0$-reach-avoid set is equal to $\mathcal{R}_0$. Herein, we also use this system for theoretical analysis.
\end{remark}

However, real-time computing is important in practice. The time delay may degrade the performance of control systems or even worse lead to safety violations \cite{xue2020over}. Thus, we would like to obtain new constraints, which are simpler and can be solved more efficiently, for computing inner-approximations. Moreover, we would like to obtain an inner-approximation such that system \eqref{system} is able to operate inside it until the target set is reached. The property of maintaining system \eqref{system} inside the computed inner-approximation is important. This will facilitate the synthesis of reach-avoid controllers online, since if system \eqref{system} leaves the computed inner-approximation online, we are no longer able to guarantee that system \eqref{system} will be able to reach the target set $\mathcal{X}_r$ safely. Besides, it is not sufficient to only ensure that the target set is reached eventually. Another practical concern is to ensure that the target hit happens within some tolerable time limit \cite{khalil2002nonlinear}. Regarding these practical demands, we in the following present a new set of constraints.

 

\begin{theorem}
\label{relax}
Given a bounded function $v(\bm{x}): \widehat{\mathcal{C}}\rightarrow \mathbb{R}$ and a factor $\lambda \in (1,\infty)$, if they satisfy 
\begin{equation}
\label{constraints_relax}
    E[v(\widehat{\bm{f}}(\bm{x},\bm{u}))]- \lambda v(\bm{x})\geq 0, \forall \bm{x}\in \widehat{\mathcal{C}}\setminus \mathcal{X}_r,\\
\end{equation}
then $\Omega=\{\bm{x}\in \mathcal{C}\mid v(\bm{x})>0\}$ is an inner-approximation of the controlled-reach-avoid set $\mathcal{R}$, i.e., $\{\bm{x}\in \mathcal{C}\mid v(\bm{x})>0\}\subseteq \mathcal{R}$, where 
the set $\widehat{\mathcal{C}}$ is a set in \eqref{sets}.
\end{theorem}

\begin{remark}
As in \cite{xue2021reach}, when the datum are polynomials, i.e., $\bm{f}(\bm{x},\bm{u})$ is polynomial over $\bm{x}$, the control set $\mathcal{U}$ and safe set $\mathcal{C}$ are bounded semi-algebraic sets, a set $\widehat{\mathcal{C}}$ satisfying Theorem \ref{relax} can be computed  via solving a semi-definite programming problem which can be solved efficiently in polynomial time via interior point methods. 
\end{remark}

Theorem \ref{relax} indicates that an inner-approximation of the $0$-reach-avoid set $\mathcal{R}_0$ can be computed via solving constraint \eqref{constraints_relax} and thus an inner-approximation of the CRS $\mathcal{R}$ can also be computed via solving it. The set of constraints \eqref{constraints_relax} can be equivalently reformulated as follows:
\begin{equation}
\label{constraints_relax_eq}
    \begin{cases}
    &E[v(\bm{f}(\bm{x},\bm{u}))]-\lambda v(\bm{x})\geq 0, \forall \bm{x}\in \mathcal{C}\setminus \mathcal{X}_r,\\
    &v(\bm{x})\leq 0, \forall \bm{x}\in \widehat{\mathcal{C}}\setminus \mathcal{C}, \\
    \end{cases}
\end{equation}
where $\lambda\in (1,\infty)$ is a given factor. 
Comparing with constraint \eqref{constraint}, constraint \eqref{constraints_relax}  abandons a bounded function $\bm{w}(\cdot):\widehat{\mathcal{C}}\rightarrow \mathbb{R}$ and thus gets simplified.

In addition, it is interesting to find that if $v(\bm{x})$ satisfies constraint \eqref{constraints_relax_eq} and $\bm{x}_0\in \Omega$, there exists $\bm{u}\in \mathcal{U}$ such that 
$v(\bm{f}(\bm{x}_0,\bm{u}))\in \Omega$. This conclusion can be obtained from the fact that 
$v(\bm{x})\leq 0, \forall \bm{x}\in \widehat{\mathcal{C}}\setminus \Omega$
and $E[v(\bm{f}(\bm{x},\bm{u}))]-\lambda v(\bm{x})\geq 0, \forall \bm{x}\in \mathcal{C}\setminus \mathcal{X}_r$, which implies that 
\[E[v(\bm{f}(\bm{x}_0,\bm{u}))]> 0.\]
This property is important since it guarantees the feasibility of maintaining system \eqref{system} inside the set $\{\bm{x}\in \mathcal{C}\mid v(\bm{x})>0\}$ before the target set $\mathcal{X}_r$ is hit. 

\begin{corollary}
\label{existence}
Let $\Omega=\{\bm{x}\in \mathcal{C}\mid v(\bm{x})>0\}$, where $v(\cdot): \widehat{\mathcal{C}}\rightarrow \mathbb{R}$ is a bounded function satisfying constraint \eqref{constraints_relax}. For any $\bm{x}_0\in \Omega$, there exists a controller $\pi\in \mathbb{U}$ such that system \eqref{system} reaches the target set $\mathcal{X}_r$ while staying inside the set $\Omega$ before the first target hitting time.
\end{corollary}

Also, constraint \eqref{constraints_relax} indicates that for any initial state $\bm{x}_0 \in \{\bm{x}\in \mathcal{C}\setminus \mathcal{X}_r\mid v(\bm{x})>0\}$ and a controller $\pi\in \mathbb{U}$ such that system \eqref{system} starting from $\bm{x}_0$ reaches the target set $\mathcal{X}_r$ while staying inside the set $\Omega$ before the first target hitting time, the first target hitting time can be upper bounded.  This addresses another practical concern, which is to ensure that the target hit happens within some tolerable time limit \cite{khalil2002nonlinear}.



\begin{corollary}
\label{bounded}
Let $\Omega=\{\bm{x} \in \mathcal{C}\mid v(\bm{x})>0\}$ and $\bm{x}_0\in \Omega\setminus \mathcal{X}_r$,  $\mathbb{U}_1$ be the set of controllers such that system \eqref{system} reaches the target set $\mathcal{X}_r$ while staying inside the set $\Omega$ before the first target hitting time, 
\[T_{\bm{x}_0}=\inf_{\pi\in \mathbb{U}_1}\inf\{k\in \mathbb{N}\mid  \bm{\psi}_{\bm{x}_0}^{\pi}(k)\in \mathcal{X}_r \wedge \bigwedge_{l=0}^k \bm{\psi}_{\bm{x}_0}^{\pi}(l)\in \Omega\}\] be the minimum time  of hitting the target set $\mathcal{X}_r$, and  
\[T_{\bm{x}_0}^{\pi}=\inf\{t\in \mathbb{N}\mid {\bm{\psi}}_{\pi}^{\bm{x}_0}(t)\in \mathcal{X}_r\}\] be the first time of reaching the target set $\mathcal{X}_r$ under the controller $\pi\in \mathbb{U}$, and $v(\cdot):\widehat{\mathcal{C}}\rightarrow \mathbb{R}$ satisfy constraint \eqref{constraints_relax} in Theorem \ref{relax}. Then, 
$T_{\bm{x}_0}\leq \log_{\lambda}\frac{M}{v(\bm{x}_0)}$. 
Moreover, 
$E^{\infty}[T_{\bm{x}_0}^{\pi}]\leq \frac{M-v(\bm{x}_0)}{(\lambda-1)v(\bm{x}_0)}$,
where $M\geq \sup_{\bm{x}\in \widehat{\mathcal{C}}} v(\bm{x})$.
\end{corollary}


If the current state of the system is $\bm{x}_0$, the computation of an inner-approximation of the CRS $\mathcal{R}$ can be addressed by solving the following constraints:
\begin{equation}
\label{constraints_relax_eq_op}
\begin{split}
\begin{cases}
     &E[v(\bm{f}(\bm{x},\bm{u}))]-\lambda v(\bm{x})\geq 0, \forall \bm{x}\in \mathcal{C}\setminus \mathcal{X}_r,\\
   &v(\bm{x})\leq 0, \forall \bm{x}\in \widehat{\mathcal{C}}\setminus \mathcal{C}, \\
    &v(\bm{x}_0)>0,
    \end{cases}
    \end{split}
\end{equation}
where $\lambda \in (1,\infty)$ is a user-defined factor. The condition $v(\bm{x}_0)>0$ is to ensure that the current state falls within the computed inner-approximation $\{\bm{x}\in \mathcal{C}\mid v(\bm{x})>0\}$.


 Given a factor $\lambda\in (1,\infty)$, when the datum are polynomials, i.e., $\bm{f}(\bm{x},\bm{u})$ is polynomial over state variables $\bm{x}$, the target set $\mathcal{X}_r$ and safe set $\mathcal{C}$ are semi-algebraic sets, and the function $v(\bm{x})$ is searched in the polynomial space, optimization \eqref{constraints_relax_eq_op} can be encoded into a semi-definite programming problem which can be solved efficiently in polynomial time via interior point methods. Assume that 
 \[\widehat{\mathcal{C}}=\{\bm{x}\in \mathbb{R}^n\mid h_0(\bm{x})\leq 0\},\]
 \[\mathcal{C}=\{\bm{x}\in \mathbb{R}^n \mid \wedge_{i=1}^l h_i(\bm{x})\leq 0\},\]
 and 
 $\mathcal{X}_r=\{\bm{x}\in \mathbb{R}^n \mid \wedge_{i=1}^k g_i(\bm{x})\leq 0\}$, the resulting semi-definite program is formulated below.  There are many powerful solvers such as SeDuMi \cite{sturm1999using} and Mosek \cite{mosek2015mosek} for solving it.
 \begin{algorithm}
\begin{equation}
\label{sos}
\small
\begin{cases}
&E[v(\bm{f}(\bm{x},\bm{\theta}))]-\lambda v(\bm{x})+\sum_{i=1}^l s_{0,j,i}(\bm{x})h_i(\bm{x})\\
&-s_{1,j}(\bm{x})g_j(\bm{x})+\sum_{i\neq j} s_{1,j,i}(\bm{x})g_i(\bm{x}) \in \sum[\bm{x}],\\
&-v(\bm{x})+s_{2,j}(\bm{x})h_0(\bm{x})-s_{3,j}(\bm{x})h_j(\bm{x})\\
&~~~~~~~~~~~~~~~~~~~~~~~+\sum_{i\neq j} s_{4,j,i}(\bm{x})h_i(\bm{x})\in \sum[\bm{x}],\\
&v(\bm{x}_0)-\epsilon\geq 0,\\
&j=1,\ldots,l,
\end{cases}
\end{equation}
where $v(\bm{x})\in \mathbb{R}[\bm{x}]$, and $s_{0,j,i}(\bm{x})$, $s_{1,j}(\bm{x})$, $s_{1,j,i}(\bm{x})$, $s_{2,j,i}(\bm{x})$, $s_{3,j}(\bm{x})$, $s_{4,j,i}(\bm{x}) \in \sum[\bm{x}]$, and $\epsilon>0$ is a user-defined positive threshold enforcing $v(\bm{x}_0)>0$.
\end{algorithm}
Otherwise, counterexample guided inductive optimization (CEGIO) approaches based on Satisfiability Modulo Theories (SMT) solvers such as dReal \cite{gao2013dreal}, can be employed to solve constraint \eqref{constraints_relax_eq_op} \cite{abate2021fossil}. 
We will investigate it in our future work.

\oomit{\begin{example}
\label{illu}
Consider an example from \cite{xue2021reach},
\begin{equation*}
    \begin{cases}
     &x(l+1)=x(l)-10^{-2}(0.5x(l)+0.5y(l)-0.5x(l)y(l))\\
     &y(l+1)=y(l)+10^{-2}(-0.5y(l)+1+u(l)),
    \end{cases}
\end{equation*}
with the safe set $\mathcal{C}=\{(x,y)^{\top}\mid x^2+y^2-1\leq 0\}$ and target set $\mathcal{X}_r=\{(x,y)^{\top}\mid 10x^2+10(y-0.5)^2-1\leq 0\}$. 

Assume that $u(l)$ for $l\in \mathbb{N}$ has the
uniform distribution over $[-1,1]$, we compute an inner-approximation of the CRS $\mathcal{R}$ via solving semi-definite program \eqref{sos}. In the computations, the degree of all of unknown polynomials in \eqref{sos} is $6$, $\lambda=1.01$, $\bm{x}_0=(0,-0.5)^{\top}$ $\epsilon=10^{-6}$ and $\widehat{\mathcal{C}}=\{(x,y)^{\top} \mid x^2+y^2-1.1\leq 0\}$. The computed inner-approximation is illustrated in Fig. \ref{illu_eps}.
\begin{figure}[htb!]
\center
\includegraphics[width=1.5in]{figures/ex_illu1_nw_new.pdf} 
\caption{green and purple curve-- $\partial \mathcal{C}$ and $\{\bm{x}\in \mathcal{C}\mid v(\bm{x})>0\}$; yellow circle -- $\partial \mathcal{X}_r$.}
\label{illu_eps}
\end{figure}
\end{example}
}

\begin{remark}
Since in practice real-time computing is important and finding feasible solutions is more efficient generally than solutions in some sense optimal, we do not add an objective function to constraint \eqref{sos}. 
\end{remark}

\subsection{Reach-avoid Controllers Synthesis} \label{subsec:RCS}
In this subsection we elucidate our method of synthesizing reach-avoid controllers based on the computed inner-approximation $\Omega=\{\bm{x}\in \mathcal{C}\mid v(\bm{x})\geq 0\}$.

Given a reference controller $\bm{u}_{ref}(\cdot): \mathbb{R}^m \rightarrow \mathbb{R}^m$, which may not be a reach-avoid controller, we aim to synthesize an action $\bm{u}$ applied to the current state $\bm{x}_{0}$ in a minimally invasive fashion via solving the following optimization problem,
\begin{equation}
    \label{controllers_s}
    \begin{split}
    &\min_{\bm{u}\in \mathcal{U}} \text{dist}(\bm{f}(\bm{x}_0,\bm{u}),\mathcal{X}_r) + p_{\omega} \| \bm{u} - \bm{u}_{ref}\|, \\
    \text{s.t.~}& v(\bm{f}(\bm{x}_0,\bm{u})) >0,
    \end{split}
\end{equation}
where $\text{dist}(\bm{f}(\bm{x}_0,\bm{u}),\mathcal{X}_r)$ denotes the distance between the target set $\mathcal{X}_r$ and the next state starting from $\bm{x}_{0}$ driven by the action $\bm{u}$, and $p_{\omega}$ is a user defined positive weighting factor.  The  reference controller is not indispensable and can be removed in our approach, which corresponds to the case $p_{\omega}$=0.

Generally, the problem of solving optimization \eqref{controllers_s} is nonlinear, which is intractable and thus time-consuming to be solved. 
However, in practice instantaneous control synthesis with real-time performance demands is preferable. Consequently, in order to synthesize a controller efficiently, we adapt a sampling based method to solve optimization \eqref{controllers_s}. The sampling method is formulated below. We 
\begin{enumerate}
\item take a set of samples $\{\bm{u}_i\}_{1\leq i\leq N}$ uniformly and independently over the set $\mathcal{U}$;
\item  evaluate $v(\bm{f}(\bm{x}_0,\bm{u}_i))$ for $i=1,\ldots,N$, and retain the set of $\{\bm{u}_i\}_{i\in \mathcal{N}}$ such that $v(\bm{f}(\bm{x}_0,\bm{u}_i))>0$, where $\mathcal{N}\subseteq \{1,\ldots,N\}$;
\item take $\bm{u}^*=\arg \min_{\bm{u}_i,i\in \mathcal{N}} \text{dist}(\bm{f}(\bm{x}_0,\bm{u}_i),\mathcal{X}_r)+ p_{\omega} \| \bm{u} - \bm{u}_{ref}\|$.
\end{enumerate}
The action $\bm{u}^*$ is applied to system \eqref{system} and drives system \eqref{system} to the new state $\bm{x}_1:=\bm{f}(\bm{x}_0,\bm{u}^*)$. Then, the above sampling optimization method is applied to this new state. This process is continued until the target set $\mathcal{X}_r$ is hit.

\section{Experiments}

In this section, we apply our approach to  a reach-avoid scenario, which attempts to synthesize a controller such that a unicycle-type autonomous vehicle can reach predefined target sets safely.  
All computations were run on a Windows System equipped with an i9-12900H 2.5GHz CPU with 16GB RAM, where the Matlab package YAMLIP \cite{lofberg2004yalmip} was employed for sum-of-squares decomposition of multivariate polynomials and Mosek was used to solve the semi-definite programming problem \eqref{sos}. 
The car is assumed as a size-free particle in the experiment. Besides this experiment, we also demonstrate our approach on several examples, which are presented in the supplemental material. For these examples, we do not learn safe sets using the supervised learning method in \cite{9341190}. Since in our approach computing inner-approximations of CRSs is crucial and constraint \eqref{constraints_relax_eq_op} is proposed for computing them from a probabilistic perspective, these examples mainly demonstrate the generality of constraint \eqref{constraints_relax_eq_op} in inner-approximating CRSs. In addition, 
the results summarized in the supplemental material also support our claim that constraint  \eqref{constraints_relax_eq_op} is simpler than \eqref{constraint} and thus can be solved more efficiently.


\begin{example}
    \label{ex_dubins_car}
    Consider a Euler version of a Dubin's car model which describes the mobility of a 2-axis vehicle:
    \begin{equation*}
        \begin{cases}
            &x(l+1)=x(l)+v(l) \cos{\theta(l)} \\
            &y(l+1)=y(l)+v(l) \sin{\theta(l)},
        \end{cases}
    \end{equation*}
where $\mathcal{U}=\{(v,\theta)^{\top} \mid v \in [0,1], \theta \in [-\pi,\pi ]\}$ is the set of control inputs (i.e., the velocity and the steering angle). 

\begin{figure}[htb!]
    \center
     \includegraphics[width=\linewidth]{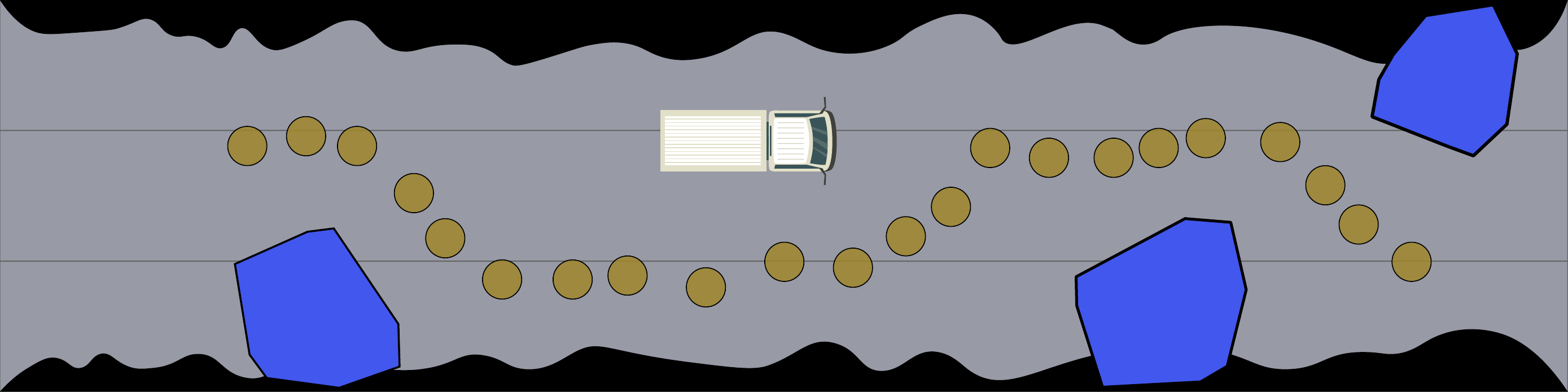}
    \caption{a $800 \times 200$ scene with 22 target sets and 4 obstacles. (Yellow regions-- target sets $\mathcal{X}_{r}$; blue, black and white regions--obstacles to be avoided)}
    \label{fig:main_scene}
\end{figure}

\begin{figure*}[t]
\center
\includegraphics[height=2.8in]{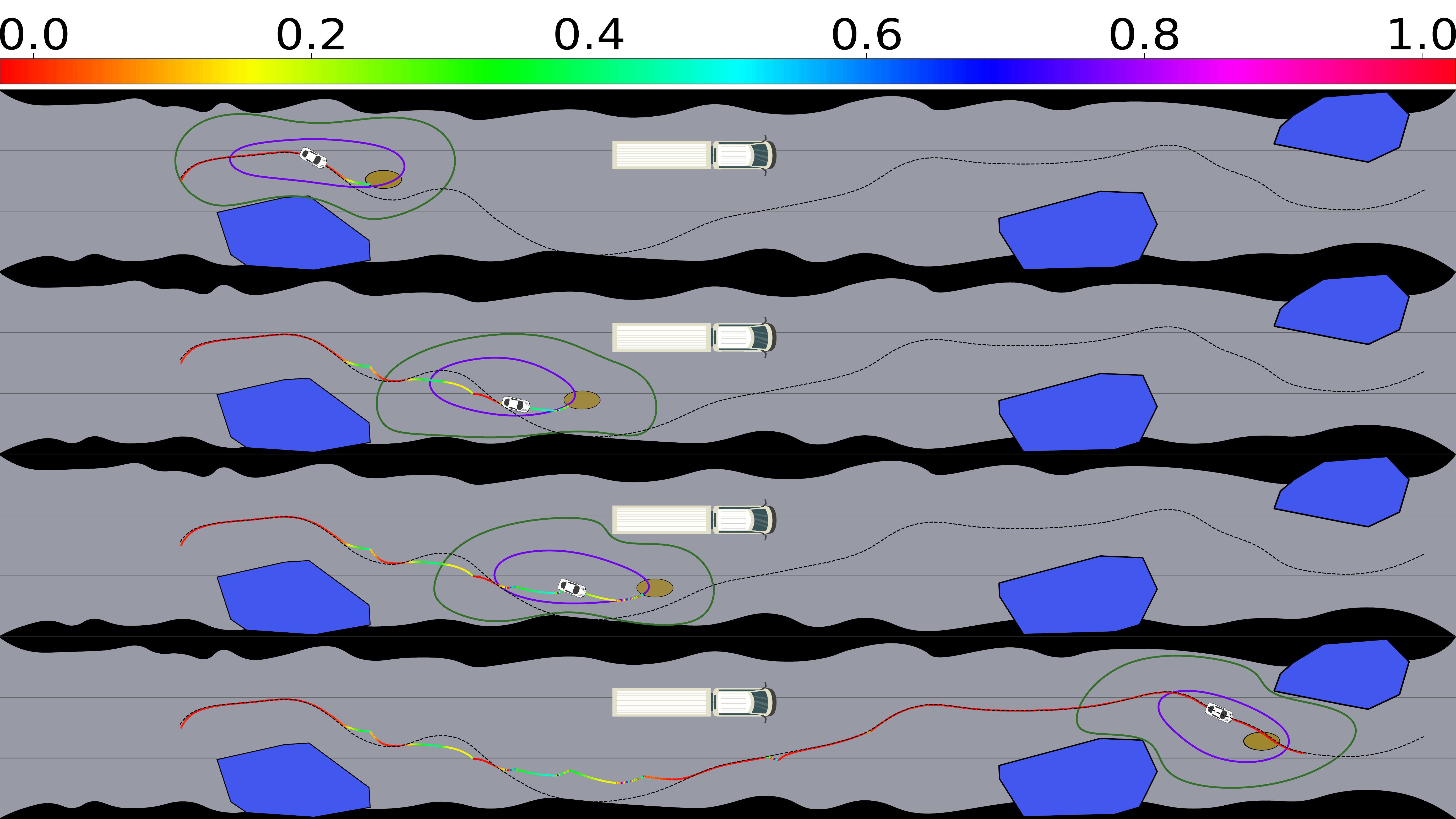} 
\caption{Scenarios on navigating to the forth, eighth, ninth, and twenty-first target sets. (Black dashed curves - trajectory generated by the reference controller; colored curves -- trajectories of the vehicle; green and purple curves -- boundaries of learned safe sets $\mathcal{C}$ and boundaries of computed inner-approximations $\{\bm{x}\in \mathcal{C}\mid v(\bm{x})>0\}$; yellow regions -- target sets $\mathcal{X}_r$)}
\label{fig:ex_main}
\end{figure*}


\begin{table*}[t]
\centering
\setlength\tabcolsep{1.5mm}
\begin{tabular}{c|c|c|c|c|c|c|c|c|c|c|c|c|c|c|c|c|c|c|c|c|c|c}
\hline
Target & 1 & 2 & 3 & 4 & 5 & 6 & 7 & 8 & 9 & 10 & 11 & 12 & 13 & 14 & 15 & 16 & 17 & 18 & 19 & 20 & 21 & 22  \\ [1ex] \hline \hline 
SVM    &1.4 &1.1 &2.6 &3.0 &1.7 &2.3 &1.6 &2.4 &2.0 &1.5 &3.2 &1.9 &2.5 &2.7 &2.3 &1.8 &2.1 &1.4 &2.1 &2.2 &1.6 &1.4     \\\hline
SDP    &3.7 &3.1 &4.5 &3.7 &4.2 &4.3 &3.9 &3.9 &4.1 &3.9 &4.9 &3.6 &4.7 &4.0 &4.2 &4.4 &3.7 &2.2 &3.5 &3.0 &3.1 &3.5
\\\hline
\end{tabular}
\caption{Computation time (seconds) in each iteration.}
\label{table1}
\end{table*}

As illustrated in the Figure \ref{fig:main_scene}, we design a scenario of size $800 \times 200$, in which a reference controller presented in \cite{fossen2014line} is introduced. This reference controller will lead the car to collide with obstacles or miss some target sets, as illustrated in Figure \ref{fig:ex_main}. 

In the experiment, we endow the set $\mathcal{U}$ with uniform probability distribution. Each iteration of reaching a target set safely begins with learning a semi-algebraic safe set (with polynomials of degree $6$) using the supervised learning method introduced in Subsection \ref{subsec:LSS}.
In addition to generating labelled data via following the way in \cite{9341190}, we also consider the point at a distance $D$ (valid scanning radius) radially outwards in the direction no obstacles are detected to be safe, while its counterpart at a further distance outwards to be unsafe.
For details, please refer to  \cite{9341190}. 
After learning a safe set, we compute an inner-approximation of the {\em CRS} by solving semi-definite program \eqref{sos}, in which the degree of all unknown polynomials is $8$.
We empirically set $p_{\omega}$, $D$, $C^+$, $C^-$ , $\lambda$ as 3, $80$, $1$, $10^{16}$, $1.01$ respectively in our experiment. 



 Figure \ref{fig:ex_main} shows the L2-norm of differences between the trajectories generated by the reference controller and the controller synthesized by our approach. Four scenarios were chosen to demonstrate the performance of controllers synthesized by our approach. They include four cases, in which the reference controller 1) fails to steer the vehicle into the target set, 2) fails to steer the vehicle into the target set and even crashes into roadside obstacles, 3) fails to steer the vehicle within the reach-avoid set, and 4) steers the vehicle within the computed reach-avoid set. The navigation to the forth, eighth and ninth target sets show that our controller is able to modify the reference controller and escort the vehicle to the predefined target sets safely. 

 The efficiency of our approach is mainly affected by computations of safe sets and inner-approximations of {\em CRSs}. Therefore, only the computation times of these two procedures are shown in Table \ref{table1}. 
In all cases, we use semi-algebraic sets to represent safe sets and CRSs, thus limiting the computations in the polynomial space. Polynomials of high degree will expand the feasible space of the resulting semi-definite program \eqref{sos} and thus increase the possibility of constructing an inner-approximation successfully. On the other hand, they will result in an increase in the amount of computation time, thus degrading efficiency. In order to overcome the predicament, more general functions beyond polynomials such as deep neural networks will be used in our computations and corresponding efficient algorithms for solving constraint \eqref{constraints_relax_eq_op} will be developed in the future. 
 

\end{example}


    

\color{black}

\section{Conclusion}
In this paper we proposed an approach for synthesizing provable reach-avoid controllers, which drive systems operating in an unknown environment to reach a desired target set safely. Based on a safe set learned from sensor measurements, the approach was built upon the computation of inner-approximations of CRSs, which provide a viability space for the system to achieve the reach-avoid objective. In order to compute such an inner-approximation efficiently, we derive a set of new constraints. Finally, we demonstrated our approach on a Dubin's car system, which is required to reach predefined target sets safely. 


 \color{red}

 \color{black}

\label{sec:con}

\bibliographystyle{named}
\bibliography{reference}
\section*{Appendix I}
\textbf{The proof of Theorem 1:}
\begin{proof}
Assume that $\bm{x}_0\in \{\bm{x}\in \mathcal{C}\mid v(\bm{x})> 0\}\setminus \mathcal{X}_r$. From the constraint \[E[v(\widehat{\bm{f}}(\bm{x},\bm{u}))]-\lambda v(\bm{x})\geq 0, \forall \bm{x}\in \widehat{\mathcal{C}},\] we have that 
\begin{equation*}
\label{inequ}
0<v(\bm{x}_0)\leq E^{\infty}[v(\bm{\psi}_{\pi}^{\bm{x}_0}(i))], \forall i\in \mathbb{N},
\end{equation*}
where $\bm{\psi}_{\pi}^{\bm{x}_0}(\cdot): \mathbb{N}\rightarrow \mathbb{R}^n$ satisfies system (6).

 Assume that for any controller $\pi\in \mathbb{U}$, $\bm{\psi}_{\pi}^{\bm{x}_0}(i)\notin \mathcal{X}_r$ for $i\in \mathbb{N}$. This implies that 
 \[\forall \pi\in \mathbb{U}. \forall i\in \mathbb{N}. \bm{\psi}_{\pi}^{\bm{x}_0}(i)\in \widehat{\mathcal{C}}\setminus \mathcal{X}_r.\]
 Consequently, we have that \[E^{\infty}[v(\bm{\psi}_{\pi}^{\bm{x}_0}(i))]\geq \lambda^i v(\bm{x}_0)>0, \forall i\in \mathbb{N},\]
and thus $\lim_{i\rightarrow \infty}E^{\infty}[v(\bm{\psi}_{\pi}^{\bm{x}_0}(i))]=\infty$, which contradicts that $v(\cdot): \widehat{\mathcal{C}}\rightarrow \mathbb{R}$ is bounded.

Consequently, there exists a controller $\pi\in \mathbb{U}$ such that 
\[\exists i\in \mathbb{N}. \bm{\psi}_{\pi}^{\bm{x}_0}(i)\in \mathcal{X}_r.\]

The proof is completed.
\end{proof}

\textbf{The proof of Corollary 1:}
\begin{proof}
    Let $\bm{x}\in \Omega$ and $\mathcal{U}_{\bm{x}}=\{\bm{u}\in \mathcal{U}\mid \widehat{\bm{f}}(\bm{x},\bm{u})>0\}$. 
    
    From the above analysis, for $\bm{x}_0\in \Omega$, there exists $\pi\in \mathbb{U}$ such that 
    \[\forall k \in \mathbb{N}. \bm{\psi}_{\bm{x}_0}^{\pi}(k)\in \Omega,\] 
where $\bm{u}(k)\in \mathcal{U}_{\bm{\psi}_{\bm{x}_0}^{\pi}(k)}$. We denote the set of such $\pi$ by $\mathbb{U}_1$. Assume that there does not exist $\pi\in \mathbb{U}_1$ such that system (6) starting from $\bm{x}_0$ reaches the target set $\mathcal{X}_r$, implying that \[\forall \pi\in \mathbb{U}_1. \forall k \in \mathbb{N}. \bm{\psi}_{\bm{x}_0}^{\pi}(k)\in \Omega\setminus \mathcal{X}_r.\]

According to constraint (7), we have that 
    \[\begin{split}
    &E[v(\widehat{\bm{f}}(\bm{x}_0,\bm{u}))]\\
    &=\int_{\mathcal{U}_{\bm{x}_0}} v(\widehat{\bm{f}}(\bm{x}_0,\bm{u})) d P(\bm{u})+\int_{\mathcal{U}\setminus \mathcal{U}_{\bm{x}_0}} v(\widehat{\bm{f}}(\bm{x}_0,\bm{u})) d P(\bm{u})\\
    &\geq \lambda v(\bm{x}_0),
    \end{split}\]
    which implies 
    $\int_{\mathcal{U}_{\bm{x}_0}} v(\widehat{\bm{f}}(\bm{x}_0,\bm{u})) d P(\bm{u})\geq \lambda v(\bm{x}_0)$ or $\int_{\mathcal{U}_{\bm{x}_0}} v(\bm{\psi}_{\pi}^{\bm{x}_0}(1)) d P(\bm{u}(0))\geq \lambda v(\bm{x}_0)$. 

    Further, we have that \[
    \begin{split}
    &\int_{\mathcal{U}_{\bm{x}_0}} E[v(\bm{\psi}_{\pi}^{\bm{x}_0}(2))] d P(\bm{u}(0)) \geq  \int_{\mathcal{U}_{\bm{x}_0}} \lambda v(\bm{\psi}_{\pi}^{\bm{x}_0}(1)) d P(\bm{u}(0))\\
    &\geq \lambda^2 v(\bm{x}_0),
    \end{split}
    \]
    which also implies that $\int_{\mathcal{U}_{\bm{x}_0}} \int_{\mathcal{U}_{\bm{\psi}_{\pi}^{\bm{x}_0}
    (1)}} v(\bm{\psi}_{\pi}^{\bm{x}_0}(2))  d P(\bm{u}(1)) d P(\bm{u}(0)) \geq \lambda^2 v(\bm{x}_0).$

By induction, we obtain that 
\begin{equation}
\label{induction}
\begin{split}
&M\geq \int_{\mathcal{U}_{\bm{x}_0}} \cdots\int_{\mathcal{U}_{\bm{\psi}_{\pi}^{\bm{x}_0}
    (k)}} v(\bm{\psi}_{\pi}^{\bm{x}_0}(k+1))  d P(\bm{u}(k)) \cdots d P(\bm{u}(0)) \\
    &\geq \lambda^{k+1} v(\bm{x}_0), \forall k\in \mathbb{N}, 
    \end{split}
\end{equation}
where $M\geq \sup_{\bm{x}\in \widehat{\mathcal{}C}} v(\bm{x})$, contradicting the fact that $v(\bm{x})$ is bounded over $\widehat{\mathcal{C}}$. 

    Therefore, there exists $\pi\in \mathbb{U}$ such that 
    \[\exists k \in \mathbb{N}. \bm{\psi}_{\bm{x}_0}^{\pi}(k)\in \mathcal{X}_r\wedge \forall l\in [0,k]\cap \mathbb{N}. \bm{\psi}_{\pi}^{\bm{x}_0}(l)\in \mathcal{C}.\] 
    The proof is completed.
\end{proof}

\textbf{The proof of Corollary 2:}
\begin{proof}
1). From inequality \eqref{induction}, we have that 
\[\lambda^{T_{\bm{x}_0}}v(\bm{x}_0)\leq M.\]
Consequently, 
\[
\begin{split}
  T_{\bm{x}_0}&\leq \log_{\lambda}\frac{M}{v(\bm{x}_0)}.
\end{split}
\]

2). From constraint (7), we have that 
\[
\begin{split}
   & E^{\infty}[v(\bm{\psi}_{\pi}^{\bm{x}_0}(i+1))]-E^{\infty}[v(\bm{\psi}_{\pi}^{\bm{x}_0}(i))]\\
   &\geq (\lambda-1)E^{\infty}[v(\bm{\psi}_{\pi}^{\bm{x}_0}(i))]\\
   &\geq (\lambda-1)E^{\infty}[v(\bm{\psi}_{\pi}^{\bm{x}_0}(i))]\cdot P^{\infty}[T_{\bm{x}_0}^{\pi}>i]\\
   &\geq (\lambda-1)v(\bm{x}_0) \cdot P^{\infty}[T_{\bm{x}_0}^{\pi}>i], \forall i\in \mathbb{N}.
\end{split}
\]
which implies that \[
\begin{split}
&E^{\infty}[v(\bm{\psi}_{\pi}^{\bm{x}_0}(l+1))]-v(\bm{x}_0)\\
&\geq (\lambda-1)v(\bm{x}_0) \cdot\sum_{i=0}^l  P^{\infty}[T_{\bm{x}_0}^{\pi}>i].
\end{split}
\]
Thus, we have that 
\[
\begin{split}E^{\infty}[T_{\bm{x}_0}^{\pi}]&=\sum_{i=0}^{\infty}  P^{\infty}[T_{\bm{x}_0}^{\pi}>i]\\
&\leq \frac{E^{\infty}[v(\bm{\psi}_{\pi}^{\bm{x}_0}(l+1))]-v(\bm{x}_0)}{ (\lambda-1)v(\bm{x}_0) }\\
&\leq \frac{M-v(\bm{x}_0)}{ (\lambda-1)v(\bm{x}_0) }.
\end{split}
\]

The proof is completed.
\end{proof}


\section*{Appendix II}

\begin{table}[h]
\centering
\begin{tabular}{c|c|c|c|c|c|c} 
\hline
\multirow{2}{*}{Degree} & \multicolumn{2}{c|}{Example \ref{illu_ex2}} & \multicolumn{2}{c|}{Example \ref{illu_ex3}} & \multicolumn{2}{c}{Example \ref{illu_ex4}}  \\ 
\cline{2-7}
& (5) & (9)  & (5) & (9)  & (5) & (9) \\ 
\hline
6                                 &3.457   &2.863                  &3.345   &2.985                  &3.853   &3.110                            \\ 
\hline
8                                 &4.049   &3.231                  &3.890   &3.611                  &7.155   &4.393                            \\ 
\hline
10                                &5.913   &3.980                  &4.762   &3.742                  &15.43  &7.222                            \\ 
\hline
12                                &11.55  &5.162                  &6.361   &4.489                  &34.65  &13.31                            \\ 
\hline
14                                &14.32  &7.064                  &9.460   &5.638                  &73.65  &25.01                            \\ 
\hline
16                                &38.11  &10.92                 &14.85  &7.297                  &175.8 &45.59                         \\ 
\hline
Average                           &12.90  &5.537                  &7.112   &4.627                  &51.76   &16.44                            \\
\hline
\end{tabular}
\caption{Computational time (seconds)}
\small {The `(5)' and `(9)' columns show computation times for solving constraints (5) and (9) using polynomials of various degrees based on semi-definite programming relaxations respectively. Note: The semi-definite program for encoding constraint (5) can be found in \cite{xue2021reach}, i.e., (12) in \cite{xue2021reach}.}
\label{table:algo_time}
\end{table}

\begin{example}
\label{illu_ex2}
Consider an example from \cite{xue2021reach},
\begin{equation*}
    \begin{cases}
     &x(l+1)=x(l)-10^{-2}(0.5x(l)+0.5y(l)-0.5x(l)y(l))\\
     &y(l+1)=y(l)+10^{-2}(-0.5y(l)+1+u(l)),
    \end{cases}
\end{equation*}
with the safe set $\mathcal{C}=\{(x,y)^{\top}\mid x^2+y^2-1\leq 0\}$ and target set $\mathcal{X}_r=\{(x,y)^{\top}\mid 10x^2+10(y-0.5)^2-1\leq 0\}$. 

Assume that $u(l)$ for $l\in \mathbb{N}$ has the
uniform distribution over $[-1,1]$, we first compute an inner-approximation of the {\em CRS} via solving semi-definite program (10), which is shown in Fig. \ref{sup_ex2}. In the computations, the degree of all of unknown polynomials in (10) is $6$, $\lambda=1.01$, $\bm{x}_0=(0,-0.5)^{\top}$, $\epsilon=10^{-6}$ and $\widehat{\mathcal{C}}=\{(x,y)^{\top} \mid x^2+y^2-1.1\leq 0\}$. Then, we synthesize a controller via the above sampling optimization method. The trajectory, which is driven by the synthesized reach-avoid controller, is also illustrated in Fig. \ref{sup_ex2}. The target hitting time is 33, which is less than $\log_{\lambda}\frac{M}{v(\bm{x}_0)}\approx 485$.
\begin{figure}[htb!]
\center
\includegraphics[width=3in]{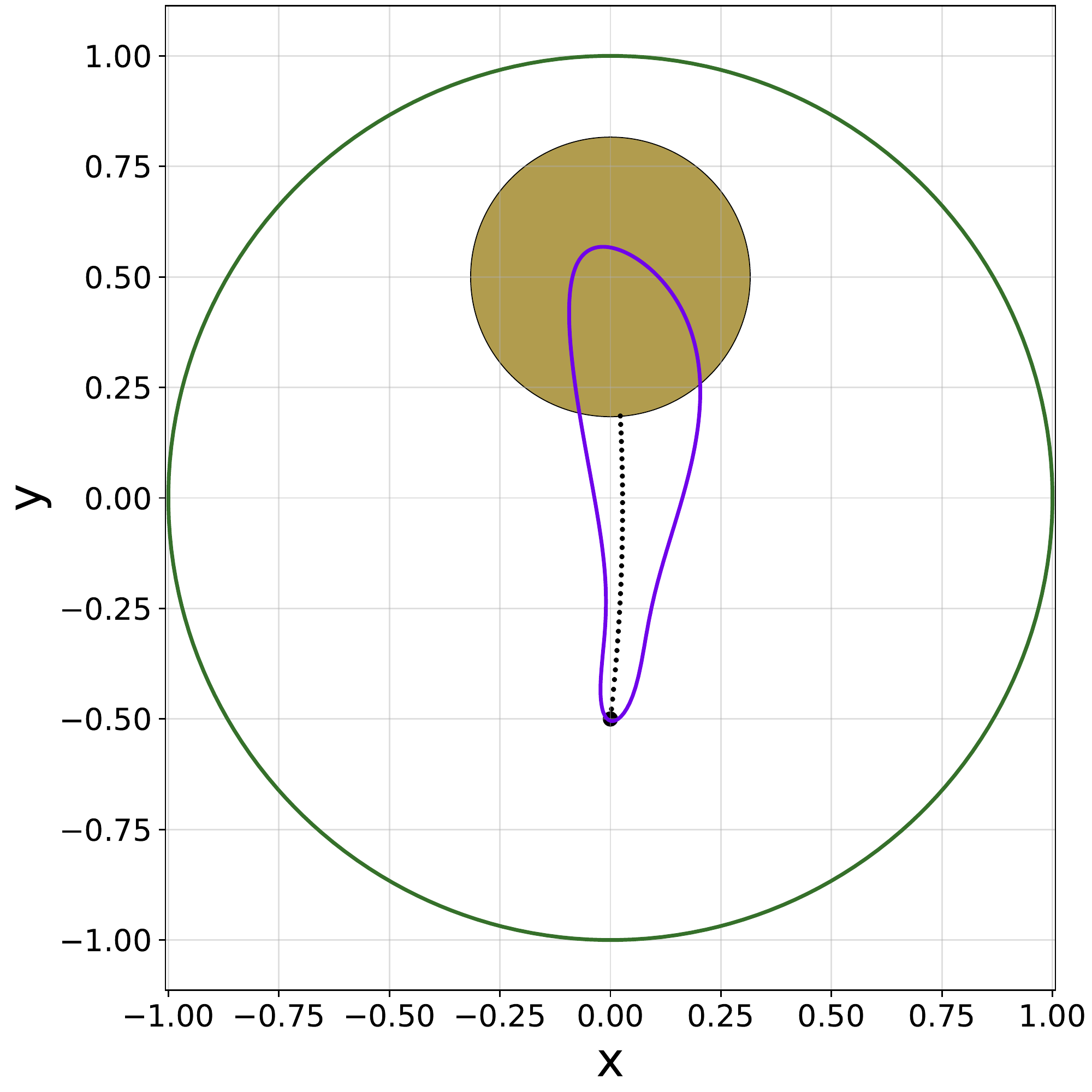} 
\caption{ black points--trajectory reaching the target set $\mathcal{X}_r$ safely; yellow region-- target set $\mathcal{X}_r$; green and purple curve--boundaries of the safe set $\mathcal{C}$ and computed inner-approximation $\{\bm{x}\in \mathcal{C}\mid v(\bm{x})>0\}$, respectively.}
\label{sup_ex2}
\end{figure}
\end{example}

\begin{example}
\label{illu_ex3}
Consider the discrete-generation predator-prey model \cite{halanay2000stability},
\begin{equation*}
    \begin{cases}
     &x(l+1)=0.5  x(l) - x(l)  y(l)\\
     &y(l+1)=-0.5  y(l) + (u(l) + 1)  x(l)  y(l),
    \end{cases}
\end{equation*}
with the safe set $\mathcal{C}=\{(x,y)^{\top}\mid x^2+y^2-1\leq 0\}$ and target set $\mathcal{X}_r=\{(x,y)^{\top}\mid 100(x^2+y^2)-1\leq 0\}$. 

Assume that $u(l)$ for $l\in \mathbb{N}$ has the
uniform distribution over $[-0.1,0.1]$, we first compute an inner-approximation of the {\em CRS} via solving semi-definite program (10), which is shown in Fig. \ref{sup_ex3}. In the computations, the degree of all of unknown polynomials in (10) is $6$, $\lambda=1.01$, $\bm{x}_0=(-0.4,-0.5)^{\top}$, $\epsilon=10^{-6}$ and $\widehat{\mathcal{C}}=\{(x,y)^{\top} \mid x^2+y^2-1.6\leq 0\}$. Then, we synthesize a controller via the above sampling optimization method. The trajectory, which is driven by the synthesized reach-avoid controller, is also illustrated in Fig. \ref{sup_ex3}. The target hitting time is 5, which is less than $\log_{\lambda}\frac{M}{v(\bm{x}_0)}\approx 187$.
\begin{figure}[htb!]
\center
\includegraphics[width=3in]{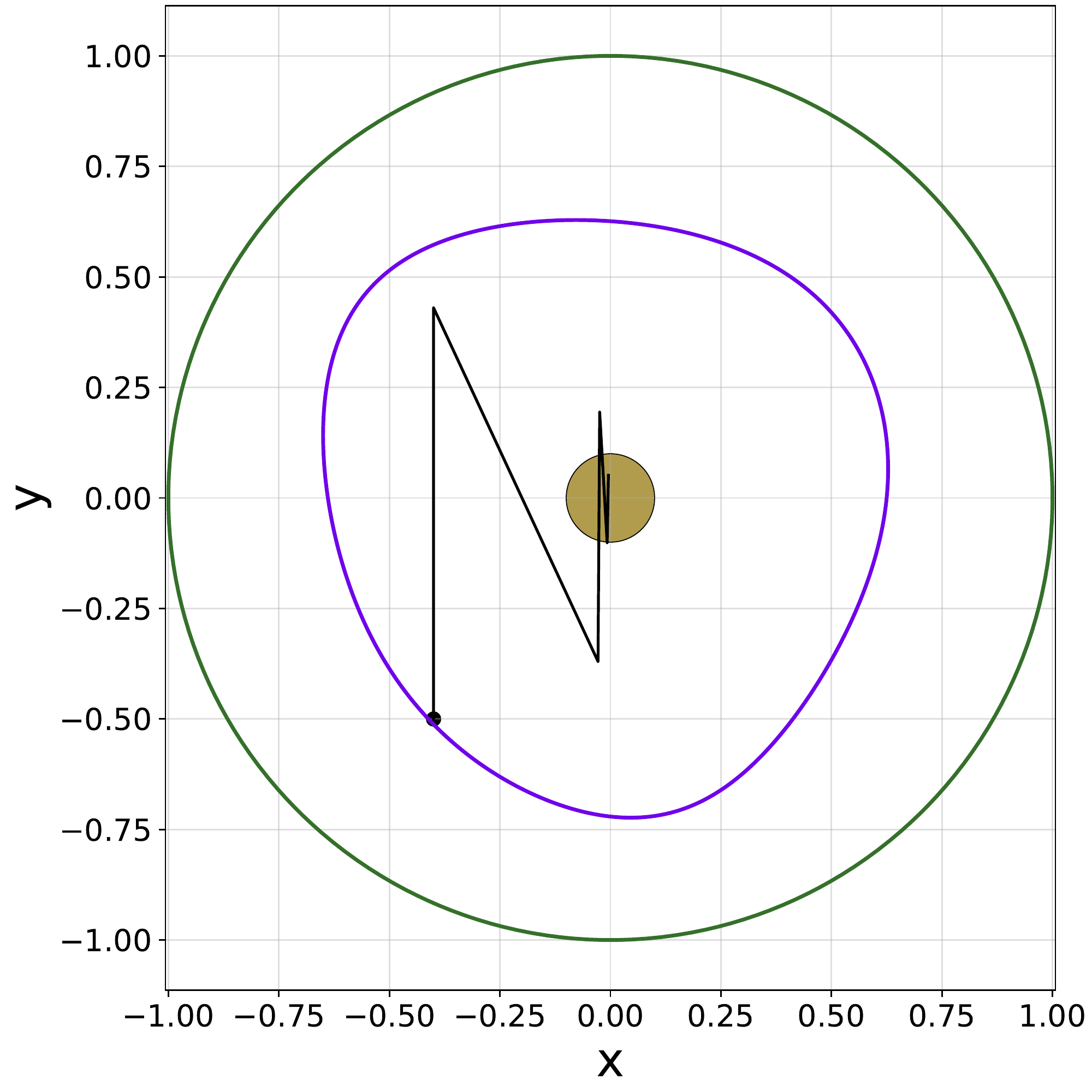} 
\caption{black points--trajectory reaching the target set $\mathcal{X}_r$ safely; yellow region-- target set $\mathcal{X}_r$; green and purple curve--boundaries of the safe set $\mathcal{C}$ and computed inner-approximation $\{\bm{x}\in \mathcal{C}\mid v(\bm{x})>0\}$, respectively.}
\label{sup_ex3}
\end{figure}

\end{example}

\begin{example}
\label{illu_ex4}

Consider an example, which is modified from \cite{tan2008stability},
\begin{equation*}
    \begin{cases}
     &x(l+1)=x(l) + 10^{-2}(y(l) + u(l))\\
     &y(l+1)=y(l) + 10^{-2}(-(1-x(l)^2)  x(l)  - y(l)),
    \end{cases}
\end{equation*}
with the safe set $\mathcal{C}=\{(x,y)^{\top}\mid x^2+y^2-1\leq 0\}$ and target set $\mathcal{X}_r=\{(x,y)^{\top}\mid 100(x^2+y^2)-1\leq 0\}$. 

\begin{figure}[htbp!]
\center
\includegraphics[width=3in]{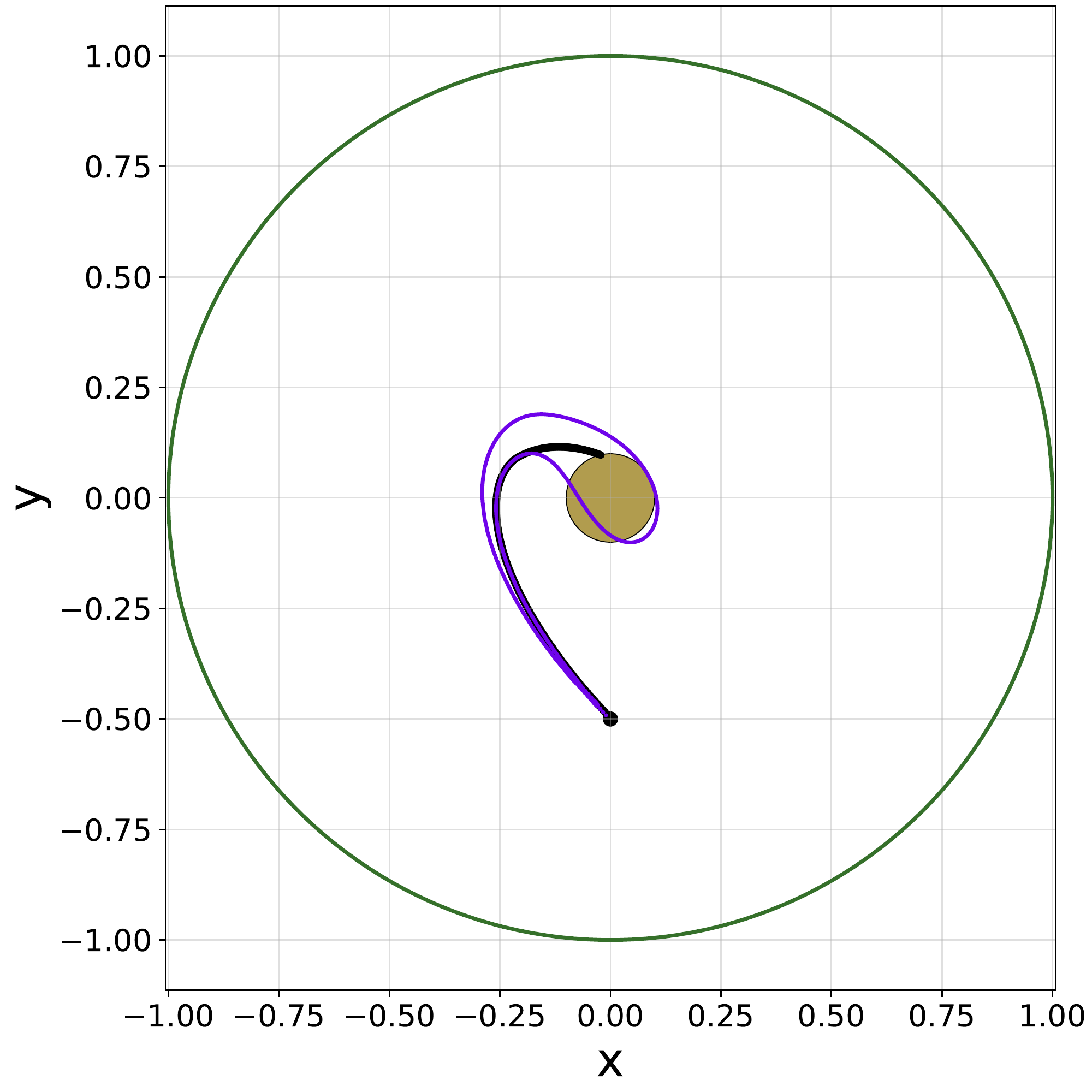} 
\caption{black points--trajectory reaching the target set $\mathcal{X}_r$ safely; yellow region-- target set $\mathcal{X}_r$; green and purple curve--boundaries of the safe set $\mathcal{C}$ and computed inner-approximation $\{\bm{x}\in \mathcal{C}\mid v(\bm{x})>0\}$, respectively.}
\label{sup_ex4}
\end{figure}

Assume that $u(l)$ for $l\in \mathbb{N}$ has the
uniform distribution over $[-0.1,0.1]$, we first compute an inner-approximation of the {\em CRS} via solving semi-definite program (10), which is shown in Fig. \ref{sup_ex4}. In the computations, the degree of all of unknown polynomials in (10) is $6$, $\lambda=1.01$, $\bm{x}_0=(0,-0.5)^{\top}$, $\epsilon=10^{-6}$ and $\widehat{\mathcal{C}}=\{(x,y)^{\top} \mid x^2+y^2-1.1\leq 0\}$. Then, we synthesize a controller via the above sampling optimization method. The trajectory, which is driven by the synthesized reach-avoid controller, is also illustrated in Fig. \ref{sup_ex4}. The target hitting time is 269, which is less than $\log_{\lambda}\frac{M}{v(\bm{x}_0)}\approx 816$.

\end{example}

\end{document}